\documentclass[conference]{IEEEtran}

\IEEEoverridecommandlockouts

\usepackage{amsmath,amsthm,amssymb,amsfonts}
\usepackage{algorithm}
\usepackage{algorithmic}
\usepackage{enumerate}
\usepackage{graphicx}
\usepackage{epstopdf}
\usepackage{url}

\newcommand{\tabincell}[2]{
\begin{tabular}{@{}#1@{}}
#2
\end{tabular}
}

\newtheorem{proposition}{Proposition}

\begin{document}


\title{Joint Resource Allocation for eICIC in Heterogeneous Networks}

\author{
\IEEEauthorblockN{Weijun Tang, Rongbin Zhang, Yuan Liu, and Suili Feng}
\IEEEauthorblockA{School of Electronic and Information Engineering\\
South China University of Technology, Guangzhou, 510641, P. R. China\\
Emails: \{tang.wj; zhang.rb\}@mail.scut.edu.cn; \{eeyliu; fengsl\}@scut.edu.cn}
\thanks{This work is supported by the National Natural Science Foundation of China under grants 61340035 and 61401159, and Science \& Technology Program of Guangzhou under grant 2014J4100246, and the SCUT-UNSW Canberra Research Collaboration Scheme.}
}

\maketitle

\vspace{-1.5cm}

\begin{abstract}
Interference coordination between high-power macros and low-power picos deeply impacts the performance of heterogeneous networks (HetNets). 
It should deal with three challenges: user association with macros and picos, the amount of almost blank subframe (ABS) that macros should reserve for picos, and resource block (RB) allocation strategy in each eNB. 
We formulate the three issues jointly for sum weighted logarithmic utility maximization while maintaining proportional fairness of users. 
A class of distributed algorithms are developed to solve the joint optimization problem. 
Our framework can be deployed for enhanced inter-cell interference coordination (eICIC) in existing LTE-A protocols. 
Extensive evaluation are performed to verify the effectiveness of our algorithms.
\end{abstract}


\IEEEpeerreviewmaketitle

\section{Introduction}
\label{Sec:Intro}

\subsection{Motivation}

Wireless data traffic has vastly grown in recent years.
The traditional cellular network can not keep pace with the data explosion.
Macro eNBs are expensive and difficult to maintain, and thus can not be deployed densely.
Hence, in Long Term Evolution-Advanced (LTE-A), a key trend of the cellular network is to increase heterogeneity through proliferation of low power nodes (e.g. pico eNBs and femto eNBs, also known as small cells).
Such heterogeneous networks (HetNets) enable a more flexible, targeted, and economical deployment of infrastructure.
Specifically, macrocells  provide a wide coverage while small cells are deployed to alleviate dead zones and traffic hot zones.
However, since picos share the spectral band with macro cells, the users within low-power picos could be severely interfered by high-power macros.
Moreover, even with a targeted deployment in high-traffic zones, most users may still receive the strongest reference signals from macro eNBs.
To address the interference management problem, eICIC has been introduced for HetNets.

\begin{figure}[t]
\begin{centering}
\includegraphics[width=85mm]{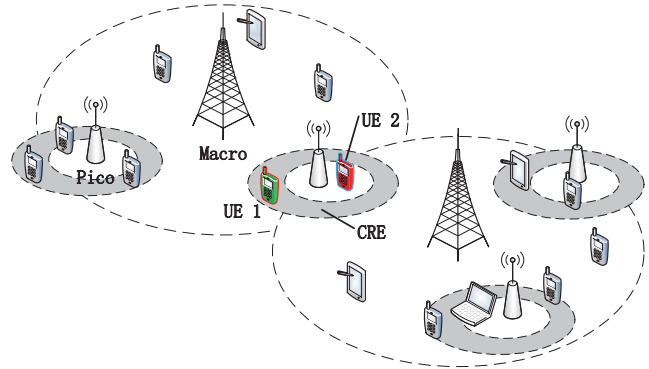}
\vspace{-0.1cm}
\caption{A typical LTE-A HetNet architecture configured with eICIC, where the shadow represent the CRE of picos.}\label{fig:Visio-system}
\vspace{-0.3cm}
\end{centering}
\end{figure}

A typical HetNet configured with eICIC as shown in Figure \ref{fig:Visio-system}, which consists of macrocells, picocells and user equipments (UEs).
In each instant, each UE is associated with one cell only, i.e. a macrocell or a picocell, bearing interferences from other cells.

There are three factors associated with the interference management problem that could restrict the network performance of HetNets.
Firstly, the unbalanced load between the two tiers.
In conventional homogeneous networks, the default UE association scheme is that maximizes the reference signal received power (RSRP) of UEs from eNBs.
It will result in unbalanced cell load in HetNets because of the widely divergent downlink transmit power of the two tiers.
In HetNets, UEs should be actively offloaded to small cells.
For this purpose, cell range extension (CRE) has been proposed for HetNets \cite{3gpp_TS36_300_2013}.
By assigning a positive bias to the RSRPs of the signals from picos, CRE extends the footprint of picos and offloads some UEs onto small cells (e.g., in Figure \ref{fig:Visio-system}, UE 1 is offloaded onto a pico).

\begin{figure}[t]
\begin{centering}
\includegraphics[width=85mm]{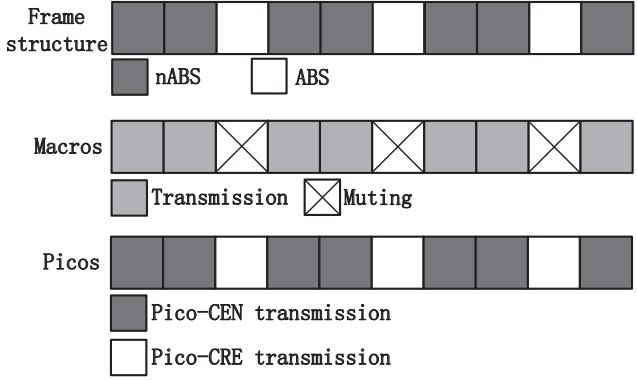}
\vspace{-0.1cm}
\caption{An illustration of ABS in HetNet with both macro- and pico-eNBs.}\label{fig:Visio-ABS}
\vspace{-0.3cm}
\end{centering}
\end{figure}

Secondly, the UEs pushed into picos would suffer degraded signal-to-interference-plus-noise-ratio (SINR), as the strongest power of the macro eNB (in terms of RSRP) now becomes to interference.
Therefore it is necessary to mitigate the interference between the two tiers. To this end, almost blank subframe (ABS) is involved in LTE-A, where macros keep ``silence'' on ABS and the offloaded UEs can be protected from macro interference. 
Figure \ref{fig:Visio-ABS} illustrates the transmission coordination using ABS.
A pico can transmit over any subframes, while a macro can only transmit over non-ABS (nABS) subframes and mutes all downlink transmissions to its UEs over ABS.
Thus, picos can transmit to their UEs over ABS with very little interference from macros.

Thirdly, how to allocate the resource blocks (RBs) among UEs in each eNB.
The channel gain of a user may change from one RB to another.
Moreover, the configuration of ABS creates two different downlink interference patterns in time domain.
Hence it is essential to study radio resource allocation for HetNets.

In this paper, we aim to design distributed algorithms for the joint problem involving above three fundamental issues, i.e.
1)How to determine user association for load balancing?
2)How to decide the optimal proportion of ABS (in time domain)? 
2)How to allocate downlink radio resources (in frequency domain)?

These three questions are coupled with each other.
To optimize the network-wide performance and realize the potential of  HetNets, it is necessary to study these problems jointly.

\subsection{Related Work}
To relieve the unbalanced load, the authors in
\cite{ye_user_2013} presented a cell association method and a distributed algorithm based on  dual decomposition method.
In \cite{wang_optimized_2011}, the cell selection problem with network-wide proportional fairness was studied by greedy heuristics.
The cell association problem was investigated by Stackelberg game in \cite{haddad_game_2013,wang_stackelberg_2013}.
The problem of optimal ABS allocation was studied in \cite{vasudevan_dynamic_2013}, the authors investigated the ABS adaptation in response to dynamic network conditions, and presented several analytical formulas.
Joint CRE and ABS configuration was studied in \cite{wang_performance_2012, wang_sensitivity_2012}.
An algorithm based on Lagrange dual method for the optimal joint CRE and ABS configuration was proposed in \cite{deb_algorithms_2013}.
The authors in \cite{jiang_resource_2012}  formulated the ABS configuration and UE association as a Nash bargaining solution (NBS).

From the review of the prior works, the joint problem of user association, ABS configuration, and radio resource allocation has not been considered yet.

\subsection{Contributions}
The main contributions and results are summarized as follows.
1)We formulate the optimization problem of user association, ABS configuration, and  radio resource allocation jointly. The goal is to maximize the sum weighted logarithmic utility of all users to maintain network-wide proportional fairness.
2)We apply the block coordinate descent (BCD) method to decouple the joint problem into three subproblems and solve each of them in an alternating manner. The proposed algorithms can be implemented distributively.

The rest of this paper is organized as follows.
Section \ref{Sec:System Model and Problem Statement} presents the system model and formulates the joint optimization problem.
Section \ref{sec:algorithm overview} studies the three components of the optimization problem, and proposes a class of algorithms.
Numerical evaluation is shown in Section \ref{Sec:Simulation}.
Finally, Section \ref{Sec:Conclusion} summarizes this paper.

\section{System Model and Problem Statement}
\label{Sec:System Model and Problem Statement}
\subsection{System Model}
Consider a HetNet consisting of $N$ UEs, $M$ macros and $P$ picos.
Denote $\mathcal{U}$, $\mathcal{M}$ and $\mathcal{P}$ as the sets of  UEs, macros and picos, respectively.
We assume a repeating ABS pattern of frames configured at the macros, and all the macros mute (or do not mute) in the same set of subframes.
$0 \leq \beta \leq 1$ is the fraction of ABS subframes
and $1-\beta$ is the fraction of nABS.
We define the UEs within the nominal coverage of the picos as ``cell-center" (CEN) UEs (e.g. UE 2 in Figure \ref{fig:Visio-system}), while the remainder of the pico UEs are defined as ``cell range-extended" (CRE) UEs (e.g. UE 1 in Figure \ref{fig:Visio-system}).
Over the nABS subframes, the UEs within CRE suffer excessive interference from the macro eNBs, while the CEN-UEs receive relatively less interference.
Hence, the UEs within the nominal coverage are more likely to be scheduled over nABS and the UEs within CRE should be protected over ABS.
Based on this observation, we consider the following policy: the UEs associated with the macros can be scheduled only during nABS; the CEN-UEs can be scheduled during nABS only, while the UEs within CRE can transmit only during ABS (see Figure \ref{fig:Visio-ABS}).
Therefore each pico can be viewed as two logical sub-eNBs: a pico-CEN and a pico-CRE. We classify the sub-eNBs into two sets: $\mathcal{P}_\textrm{CEN}$ and $\mathcal{P}_\textrm{CRE}$.
We assume that all eNBs have full buffers and fixed transmit power.
The configuration of ABS creates two different downlink interference patterns.
The SINR on RB-$ r $ of UE-$u$ from pico-$b$ can be written as
\begin{align}
&\textrm{SINR}_{ubr}= \nonumber\\
&\begin{cases}
\dfrac{P_{br} G_{ubr}}{\sum\limits_{k \in \mathcal{M}} P_{kr} G_{ukr} + \sum\limits_{k \in \mathcal{P}, k \neq b} P_{kr} G_{ukr}   + N_0}, & \textrm{for nABS},\\
\dfrac{P_{br} G_{ubr}}{ \sum\limits_{k \in \mathcal{P}, k \neq b} P_{kr} G_{ukr}   + N_0}, & \textrm{for ABS},
\end{cases}
\end{align}
where $ P_{br} $ is the transmission power assigned on RB-$r$ by eNB-$ b $.
In this paper, we assume that $\{P_{br}\}$ are uniform in each tier.
$ G_{ubr} $ is the channel gain between eNB-$b$ and UE-$u$ on RB-$r$.
During ABS subframes, all macros keep silent and so the interference is only from other picos.
However, there is interference from all other eNBs during nABS.
Instead, the SINR expression of a macro UE-$u$  is
\begin{equation}
\textrm{SINR}_{ubr} = \dfrac{P_{br} G_{ubr}}{\sum\limits_{k \in \mathcal{M}, k \neq b} P_{kr} G_{ukr} + \sum\limits_{k \in \mathcal{P}} P_{kr} G_{ukr}   + N_0}.
\end{equation}

Hence $ R_{ubr}^\textrm{nABS} $ and $ R_{ubr}^\textrm{ABS} $ denote the data rate of UE-$u$ scheduled by eNB-$b$ on RB-$r$ for nABS and ABS, respectively:
\begin{align}
\begin{split}
\label{eqn:DataRate1}
R_{ubr}^\textrm{nABS}=
\begin{cases}
\log_2 ( 1 + \textrm{SINR}_{ubr}),&\textrm{if }b \in \mathcal{M} \cup \mathcal{P}_\textrm{CEN},\\
0,&\textrm{if } b \in \mathcal{P}_\textrm{CRE},
\end{cases}
\end{split}
\\
\begin{split}
\label{eqn:DataRate2}
R_{ubr}^\textrm{ABS}=
\begin{cases}
0, & \textrm{if } b \in \mathcal{M} \cup \mathcal{P}_\textrm{CEN},\\
\log_2 ( 1 + \textrm{SINR}_{ubr}), & \textrm{if } b \in \mathcal{P}_\textrm{CRE}.
\end{cases}
\end{split}
\end{align}

\subsection{Problem Statement}
\label{Sec:Statement}
Let $S_{ub}$ be the indicator to the UE association, i.e., $S_{ub} = 1$ means that UE-$u$ is associated with eNB-$b$, and $S_{ub} = 0$ otherwise.  A UE can only associate with one eNB, and thus the UE association constraint is
\begin{equation}
\label{eqn: Ass-Con}
\sum_{b \in \mathcal{B}} S_{ub} = 1, \forall S_{ub} \in \{0,1\}, \forall u \in \mathcal{U},
\end{equation}
where $\mathcal{B} = \mathcal{M} \cup \mathcal{P}_\textrm{CEN} \cup \mathcal{P}_\textrm{CRE}$ is the collection of all logical eNBs.

Denote $x_{ubr}$ as the proportion of RB-$r$ allocated to UE-$u$ by eNB-$b$ during ABS.
Denote $y_{ubr}$ as the proportion of RB-$r$ allocated to UE-$u$ by eNB-$b$ during nABS.
An eNB can only allocate its RBs to the UEs associated with it. Hence we have the RB allocation constraints:
\begin{align}
&\sum_{u \in \mathcal{U}, } x_{ubr} = \left( 1-\beta \right),  \quad \forall b \in \mathcal{B}, \forall r \in \mathcal{R}, \label{eqn: Res-nonABS-Con}\\
&\sum_{u \in \mathcal{U}} y_{ubr} = \beta, \quad \forall b \in \mathcal{B}, \forall r \in \mathcal{R}, \label{eqn: Res-ABS-Con}\\
& 0 \leq x_{ubr} \leq S_{ub}, \quad \forall u \in \mathcal{U}, \forall b \in \mathcal{B}, \forall r \in \mathcal{R} \label{ineqn:ABS-Res-Ass},\\
& 0 \leq y_{ubr} \leq S_{ub}, \quad \forall u \in \mathcal{U}, \forall b \in \mathcal{B}, \forall r \in \mathcal{R} \label{ineqn:nABS-Res-Ass}.
\end{align}

It is well known that logarithmic utility objective makes a good balance between system throughput and user fairness.
Thus we optimize the sum weighted logarithmic utility.
Then the considered joint optimization problem can be formulated as

\begin{equation}
\label{OPT-ORG}
\begin{split}
\max\limits_{\big\{
\stackrel{S_{ub}, \beta,}{x_{ubr}, y_{ubr}}
\big\}} &\sum_{b \in \mathcal{B}} \sum_{u \in \mathcal{U}} S_{ub} \omega_u \log \Big[ \sum_{r \in \mathcal{R}} \big( R_{ubr}^\textrm{nABS} x_{ubr} + R_{ubr}^\textrm{ABS} y_{ubr} \big) \Big]  \\
\textrm{s.t.} & \quad \eqref{eqn: Ass-Con} - \eqref{ineqn:nABS-Res-Ass},
\end{split}
\end{equation}
where weights $\{\omega_u\}$ represent the grade of service among UEs.
\begin{proposition}
The joint problem is NP-hard even with a single macro and a single pico-CRE.
\end{proposition}


\section{Proposed Solution}
\label{sec:algorithm overview}
Since the computational hardness of the joint problem, we seek to find a good suboptimal solution. 
We use the BCD method \cite{razaviyayn_unified_2013} in this paper, which is also known as the Gauss-Seidel method.
The BCD method is widely used for optimizing a function of several block variables.
At each iteration of this method, a single block of variables is optimized, while the remaining variables are fixed.
In this part, we decompose the optimization variables into three blocks: $\{S_{ub}\}$, $\beta$ and $\{x_{ubr}, y_{ubr}\}$, which relate with the three sub-problems respectively: the UE association, the ABS allocation, and the radio resource allocation.
We will show in simulations that the proposed algorithms converge fast.
\subsection{Optimizing $\{x_{ubr}, y_{ubr}\}$ for Given $\{S_{ub}\}$ and $\beta$}
\label{Sec:Scheduling}
In this subsection, we study the radio resource allocation, given the ABS allocation and the UE association are fixed.
Considering \eqref{eqn:DataRate1} and  \eqref{eqn:DataRate2}, we can decompose the optimization problem into two sets of  subproblems as follows \\
$\forall b \in \mathcal{M} \cup \mathcal{P}_\textrm{CEN}$:
\begin{align}
\label{OPT-CEN-Scheduling}
\max_{\{x_{ubr}, y_{ubr}\}} & \sum\limits_{u \in \mathcal{U}_b} \omega_u \log\Big(\sum\limits_r R_{ubr}^\textrm{nABS} x_{ubr}\Big) \nonumber \\
\textrm{s.t.} & \sum\limits_{u \in \mathcal{U}_b} x_{ubr} = (1-\beta), \quad \forall r \in \mathcal{R}  \\
& 0 \leq x_{ubr} \leq 1, \quad \forall r \in \mathcal{R}, \forall u \in \mathcal{U}_b, \nonumber
\end{align}
$\forall b \in \mathcal{P}_\textrm{CRE}$:
\begin{align}
\label{OPT-CRE-Scheduling}
\max_{\{x_{ubr}, y_{ubr}\}} & \sum\limits_{u \in \mathcal{U}_b} \omega_u \log\Big(\sum\limits_r R_{ubr}^\textrm{ABS} y_{ubr}\Big) \nonumber \\
\textrm{s.t.} & \sum\limits_{u \in \mathcal{U}_b} y_{ubr} = \beta, \quad \forall r \in \mathcal{R}    \\
& 0 \leq y_{ubr} \leq 1, \quad \forall r \in \mathcal{R}, \forall u \in \mathcal{U}_b.\nonumber
\end{align}

One can see that the above sub-problems \eqref{OPT-CEN-Scheduling} and \eqref{OPT-CRE-Scheduling} are convex and mutually independent, and can be solved by the proportional fairness (PF) scheduling algorithm \cite{hou_self-organized_2012} with distributed manner in each eNB.
The details are omitted here.

\subsection{Optimizing $\{S_{ub}\}$ for Given $\{x_{ubr}, y_{ubr}\}$ and $\beta$}
\label{Sec:association}
In this subsection, we discuss the UE association, given the solutions to the ABS allocation and the radio resource allocation.
Here we consider the case in long tern such that the data rate for a UE on different RBs can be viewed as identical.
From this perspective, we provide the close-form optimal solution to the radio resource allocation problem in the following proposition.
\begin{proposition}
Given the UE association and the ABS allocation, the optimal radio resource allocation is given by:
\begin{align}
x_{ubr}^* &= \frac{\omega_u \left(1-\beta\right)}{\sum\limits_{k}S_{kb}\omega_k}, \label{eqn:Res-allo-nABS}\\
y_{ubr}^* &= \frac{\omega_u \beta}{\sum\limits_{k}S_{kb}\omega_k}. \label{eqn:Res-allo-ABS}
\end{align}
\end{proposition}

Therefore, plugging \eqref{eqn:Res-allo-nABS} and  \eqref{eqn:Res-allo-ABS} in \eqref{OPT-ORG}, the UE association problem can be written as:
\begin{align}
\label{eqn: Ass-problem}
\max\limits_{\{S_{ub}\}} \quad&\sum\limits_{b \in \mathcal{B}} \sum\limits_{u \in \mathcal{U}} S_{ub} \omega_u \log\bigg(\frac{\omega_u R_{ub}}{\sum\limits_k S_{kb} \omega_k}\bigg) \nonumber \\
\textrm{s.t.} \quad&\sum\limits_b S_{ub} = 1, \forall u \in \mathcal{U}\\
&S_{ub} \in \{0,1\}, \forall u \in \mathcal{U}, \forall b \in \mathcal{B}, \nonumber
\end{align}
where $R_{ub} = n [R_{ub}^\textrm{nABS} (1-\beta) + R_{ub}^\textrm{ABS} \beta ]$, and $n$ is the number of RBs.
The subproblem \eqref{eqn: Ass-problem} is NP-hard.
We relax the binary constraint on $\{S_{ub}\}$ and treat the sub-problem as a nonlinear programming (NLP):

\begin{align}
\label{eqn:NLP}
\max\limits_{\{S_{ub}\}} \quad&\sum\limits_{b \in \mathcal{B}} \sum\limits_{u \in \mathcal{U}} S_{ub} \omega_u \log\bigg(\frac{\omega_u R_{ub}}{\sum\limits_k S_{kb} \omega_k}\bigg) \nonumber \\
\textrm{s.t.} \quad&\sum\limits_b S_{ub} = 1, \quad \forall u \in \mathcal{U}\\
&0 \leq S_{ub} \leq 1, \quad \forall u \in \mathcal{U}, \forall b \in \mathcal{B}. \nonumber
\end{align}
\begin{proposition}
The relaxed UE association problem in \eqref{eqn:NLP} is convex.
\end{proposition}

The problem in \eqref{eqn:NLP} can be solved by standard techniques of convex optimization under central processing. %
Note that ignoring the binary constraint means that UEs can receive radio resources from more than one eNB, which violates the primal constraint of single eNB association but acts as a performance upper bound in our simulations.

%

The NLP approach cannot be deployed in distributed manner.
In the rest of this subsection, we propose a distributed UE association strategy which enforces each UE to associate with one eNB only.
It is based on the gradient descent method, which is the basic approach to unconstrained optimization problems.

Let $f$ be the objective function of \eqref{eqn:NLP}, the gradient with respect to $S_{ub}$ is
\begin{equation*}
\frac{\partial f}{\partial S_{ub}} = \omega_u \Bigg(\log \bigg(\frac{R_{ub}}{\sum\limits_{k \in \mathcal{U}} S_{kb} \omega_k}\bigg) -1 \Bigg),\forall u \in \mathcal{U}, \forall b \in \mathcal{B}.
\end{equation*}
The fraction $\{\partial f/\partial S_{ub}\}$ can be viewed as the marginal utility when UE-$u$ is associated with eNB-$b$.
Thus, we propose the following principle for choosing the best serving base station for UE-$u$:
\begin{equation}
\label{eqn:Best-BS}
S_{ub}=
\begin{cases}
1, \textrm{if } b = \arg_{k} \max \Big\{\frac{\partial f}{\partial S_{uk}}\Big\},\\
0, \textrm{otherwise}.
\end{cases}
\end{equation}

Denote $\Omega_b$ as the sum of the weights of the UEs associated with eNB-$b$, i.e. $\Omega_b = \sum\limits_{u \in \mathcal{U}} S_{ub} \omega_u$.
We propose Algorithm \ref{Heuristic Association Strategy} for the UE association problem, which can be implemented in distributed manner.
\begin{algorithm}
\caption{Heuristic Association Strategy: Algorithm for Solving the UE Association Problem}
\label{Heuristic Association Strategy}
\begin{algorithmic}[1]
\STATE \textbf{UE's Action}. For all $u \in \mathcal{U}$, perform the following steps.
\begin{enumerate}[(i)]
\item Measure the data rates $\{R_{ub}^\textrm{nABS}, R_{ub}^\textrm{ABS}\}$ from all eNBs and send them to the serving eNB.
\end{enumerate}
\STATE \textbf{eNB's Action}. For all $b \in \mathcal{B}$, perform the following steps.
\begin{enumerate}[(i)]
\item Broadcast $\Omega_b$ to other eNBs, and receive the data rates reported from UEs.
\item Determine the best serving eNB according to  \eqref{eqn:Best-BS} for each associated UE.
\item Do the handover of the UEs whose best serving eNB is changed.
\end{enumerate}
\end{algorithmic}
\end{algorithm}
The value $\{\Omega_b\}$ serves as a message between eNBs in the system.
It represents the load of each eNB.
By broadcasting $\{\Omega_b\}$, each eNB has the load information of others.
With the help of these values and the data rates reported from UEs, the eNBs can offload UEs to the light-loaded neighbor cells.
To further reduce the amount of broadcasted information, the eNBs can exchange the messages only with its neighbors in practical deployments.


\subsection{Optimizing $\beta$ for Given $\{S_{ub}, x_{ubr}, y_{ubr}\}$}
\label{Sec:ABS}
In this subsection, the optimal ABS proportion is derived.
Given the solution of $\{S_{ub}, x_{ubr}, y_{ubr}\}$, the joint problem can be rewritten as
\begin{align}
\max\limits_{\beta} \quad&
\sum\limits_{b \in \mathcal{M} \cup \mathcal{P}_\textrm{CEN}} \sum\limits_{u \in \mathcal{U}_b} \omega_u \log \bigg(\frac{n \omega_u R_{ub}^\textrm{nABS}(1-\beta)}{\sum_{k \in \mathcal{U}_b} \omega_k} \bigg) \nonumber \\
& \qquad \qquad +\sum\limits_{b \in \mathcal{P}_\textrm{CRE}} \sum\limits_{u \in \mathcal{U}_b} \omega_u \log \bigg(\frac{n \omega_u R_{ub}^\textrm{ABS}\beta}{\sum_{k \in \mathcal{U}_b} \omega_k}\bigg)
\\
\textrm{s.t.} \quad& \beta \in [0,1]. \nonumber
\end{align}

Using the optimization principles, let the gradient of the objective function equal zero then we have
\begin{equation}
-\sum\limits_{b \in \mathcal{M} \cup \mathcal{P}_{CEN}} \sum\limits_{u \in \mathcal{U}_b} \frac{\omega_u}{(1-\beta)} + \sum\limits_{b \in \mathcal{P}_{CRE}} \sum\limits_{u \in \mathcal{U}_b} \frac{\omega_u}{\beta} = 0,
\end{equation}
and the optimal ABS configuration is
\begin{equation}
\label{eq: optimal ABS}
\beta^* = \frac{\sum\limits_{b \in \mathcal{P}_{CRE}} \sum\limits_{u \in \mathcal{U}_b} \omega_u}{\sum\limits_{u \in \mathcal{U}}\omega_u}.
\end{equation}
In other words, the optimal ABS rate tracks the fractional sum of the UEs' weights in the pico-CREs. This is because the ABS resources can only be allocated to the UEs associated within the pioc-CREs. Note that each eNB can calculate the optimal ABS proportion in distributed manner by broadcasting the sum of the weights of the associated UEs to others.

\section{Performance Evaluation}
\label{Sec:Simulation}
We consider a two-tier LTE HetNet with co-channel deployment of macros and picos.
The network topology consists of a standard hexagonal grid of three-sector macro-eNBs and a set of picos with omni-directional antennas.
The transmit power of the two tiers are 46dBm and 30dBm respectively, which are distributed on each RB uniformly.
A dynamic system-level simulator is used, including explicit modeling of RB scheduling, link adaptation and channel quality indicator (CQI) feedback.
For the simulations, two picos are deployed in each macro-sector.
The location of the picos in one macro-sector are uniformly and independently distributed in space.
We consider the area within 40m radius of each pico-eNB as a hotspot.
According to the 3GPP simulation guideline \cite{3gpp_TR36_814_2010}, 2/3 of UEs are inside the hotspots and the remaining are uniformly distributed within the macro area.
In the propagation environment modeling, we use the path model in \cite{3gpp_TS36_942_2012} and lognormal shadowing with a standard deviation $\sigma = 10$dB.
WINNER II Channel Models \cite{hentila_matlab_2007} is used for channel fast fading and the thermal noise power is -174dBm/Hz.
The default simulation parameters are summarized in Table \ref{Table:simulation parameters}.
\begin{table}[t]
\renewcommand{\arraystretch}{1.3}
\caption{Summary of Simulation Parameters}
\label{Table:simulation parameters}
\centering
\begin{tabular}{|c|c|}
\hline
\bfseries Parameter & \bfseries Setting\\
\hline
\hline
Network layout & \tabincell{c}{500m macro-tier inter-site distance,\\ three sectors per macro, 2 picos \\ per macro-sector } \\
\hline
Cell layout & \tabincell{c}{7 macrocells (21 macro sectors), \\ wrap-around} \\
\hline
Number of UEs and weights & \tabincell{c}{1260 in the whole topology; uniform} \\
\hline
UE placement & \tabincell{c}{20 UEs inside the area within 40m\\ radius of each pico (the hotspot); \\the rest UEs are uniformly \\distributed within the macro area} \\
\hline
Transmit power & \tabincell{c}{Macro tier: 46dBm; pico tier: 30dBm} \\
\hline
Frame duration & 10ms (10 sub-frames)\\
\hline
Carrier and Bandwidth & \tabincell{c}{2.14GHz; 20MHz}\\
\hline
Path loss & \tabincell{c}{TS 36.942 \cite{3gpp_TS36_942_2012}, urban}\\
\hline
Shadow fading & Lognormal, standard deviation: 10dB\\
\hline
Fast fading & \tabincell{c}{WINNER II Channel Models \cite{hentila_matlab_2007}}\\
\hline
Thermal noise power & \tabincell{c}{-174dBm/Hz}\\
\hline
UE speed & 0/5 km/h\\
\hline
Traffic model & Full buffer\\
\hline
\end{tabular}
\end{table}

We investigate the situation with motionless UEs.
Denote (A,B,C) as the joint scheme for the joint problem, where A denotes the algorithm for the UE association, B denotes the algorithm for the ABS allocation, and C denotes the algorithm for the radio resource allocation. 
In the traversal experiment of joint CRE and ABS configuration with PF scheduling algorithm, we find that the case of CRE $= 18$dB and $\beta = 0.4$ provides the best system performance.
In the following, (CRE, $\beta$, PF) = ($18$, $0.4$, PF) and the conventional scheme (i.e. max-RSRP association policy without ABS) serve as the benchmarks.

\begin{figure}[t]
\begin{centering}
\includegraphics[width=85mm]{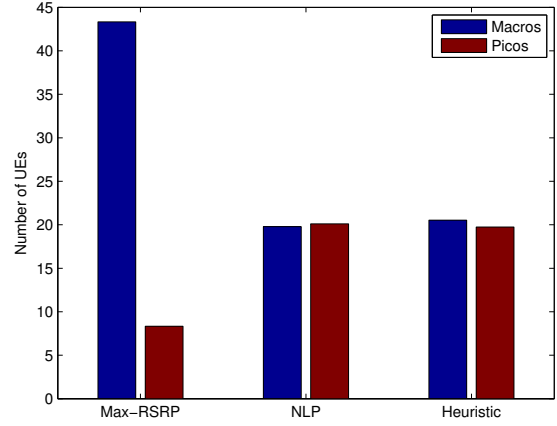}
\vspace{-0.1cm}
\caption{Average number of UEs per eNB.}\label{fig:number_of_UE_per_tier}
\vspace{-0.3cm}
\end{centering}
\end{figure}
Figure \ref{fig:number_of_UE_per_tier} compares the average number of UEs per eNB among different association schemes.
As expected, the conventional max-RSRP scheme results in very unbalanced loads: the macros are over-loaded, while the picos serve far fewer UEs.
In other two schemes, the number of UEs in the two tiers are more balanced.
The heuristic association algorithm provides close performances to the NLP solution which is optimal in the UE association.

\begin{figure}[t]
\begin{centering}
\includegraphics[width=85mm]{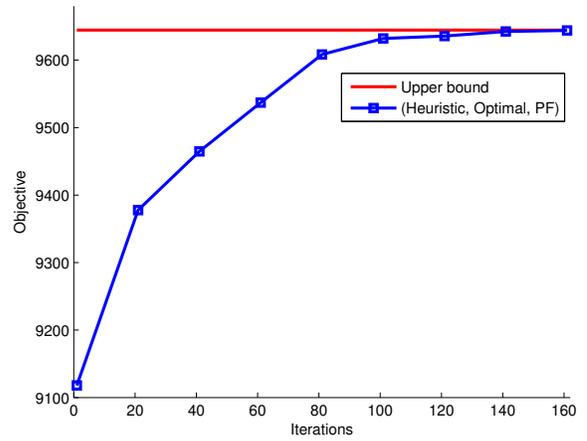}
\vspace{-0.1cm}
\caption{The convergence of different schemes.}\label{fig:convergence}
\vspace{-0.3cm}
\end{centering}
\end{figure}

Figure \ref{fig:convergence} shows the curves of iteration for our scheme.
One can see that, the proposed schemes (Heuristic, Optimal, PF), where Optimal denotes the optimal ABS ratio \eqref{eq: optimal ABS}, is convergent and provides a very close performance to the upper bound.

\begin{figure}[t]
\begin{centering}
\includegraphics[width=85mm]{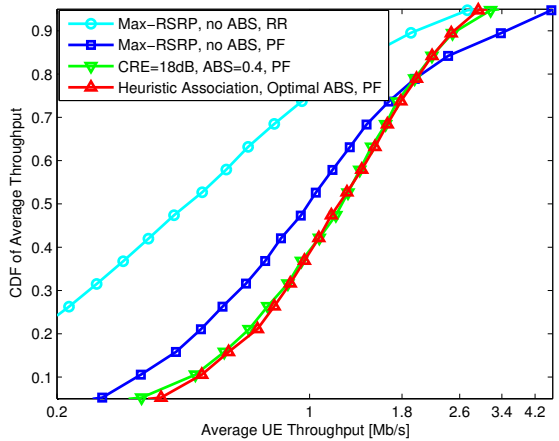}
\vspace{-0.1cm}
\caption{The CDFs of average throughput (motionless UEs).}\label{fig:CDF_average_throughput_motionless}
\end{centering}
\vspace{-0.3cm}
\end{figure}
Figure \ref{fig:CDF_average_throughput_motionless} shows the cumulative distribution function (CDF) of UE average throughput in the whole network.
Round Robin (RR) is one of the conventional scheduling schemes in LTE-A.
About 80\% of UEs access to a significantly improvement with our proposed algorithms against the max-RSRP scheme.
The proposed scheme (Heuristic, Optimal, PF) even outperforms the scheme ($18$, $0.4$, PF).
It is because that the scheme (Heuristic, Optimal, PF) computes with which eNBs the UEs should be associated, while the schemes ($18$, $0.4$, PF) can only balance the load by adapting the biases.
The former scheme is more accurate and hence provides a better performance.

Table \ref{Table:fairness} shows the fairness of  throughput among UEs for different schemes.
We consider the Jain's fairness index which can be expressed as
\begin{equation}
\frac{(\sum_{u=1}^{N} T_u)^2}{N\sum_{u=1}^{N}T_u^2},
\end{equation}
where $T_u$ is the throughput of UE-$u$.
Our proposed algorithms improve the service fairness significantly.
\begin{table}[t]
\renewcommand{\arraystretch}{1.3}
\caption{Fairness for Different Schemes (Motionless UEs)}
\label{Table:fairness}
\centering
\begin{tabular}{|c|c|}
\hline
Scheme & Fairness index\\
\hline
\hline
max-RSRP, no ABS, RR& 0.353\\
\hline
max-RSRP, no ABS, PF& 0.449\\
\hline
CRE = 18dB, ABS = 0.4, PF & 0.671\\
\hline
Heuristic algorithm, Optimal ABS, PF & 0.693\\
\hline
\end{tabular}
\end{table}

\begin{figure}[t]
\begin{centering}
\includegraphics[width=85mm]{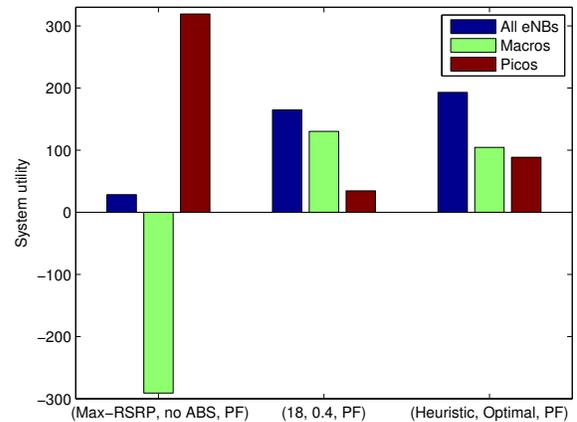}
\vspace{-0.1cm}
\caption{System utility of different tiers (motionless UEs)}\label{fig:system_utility_per_tier}
\vspace{-0.3cm}
\end{centering}
\end{figure}

Figure \ref{fig:system_utility_per_tier} shows the system utility for different schemes.
One can see that our proposed algorithms outperform the (max-RSRP, no ABS, PF) scheme in network-wide system utility.
It is because that the proposed algorithms enhance the performance of the edge UEs, as shown in Figure \ref{fig:CDF_average_throughput_motionless}.

%
%

\section{Conclusion}
\label{Sec:Conclusion}
In this paper, we consider a joint problem of RB scheduling, UE association and ABS allocation.
The goal is to maximize the weighted sum logarithmic utility of all UEs.
We solve this joint problem by using the BCD method and developed a class of algorithms.
The simulation results demonstrated that our proposed algorithms improve the network-wide resource utilization and mitigate the over-load of macro eNBs.

\bibliographystyle{IEEEtran}
\bibliography{IEEEabrv,Zotero}

\end{document}